\documentclass[sigconf]{acmart}

\usepackage{times}
\usepackage{ctable}
\usepackage{multirow}
\usepackage{balance}
\usepackage{booktabs}
\usepackage{hyphsubst}
\usepackage{graphicx}
\usepackage{amsmath}
\usepackage{url}
\usepackage{breakurl}
\usepackage{hhline}
\usepackage[caption = false]{subfig}

\newcounter{lizcounter}
\DeclareRobustCommand{\liz}[1]{\textbf{/* #1 (liz) */}\stepcounter{lizcounter}\typeout{LaTeX Warning: liz comment \thelizcounter: #1 (line \the\inputlineno)}}

\newcounter{dominikcounter}
\DeclareRobustCommand{\dominik}[1]{\textbf{/* #1 (dominik) */}\stepcounter{dominikcounter}\typeout{LaTeX Warning: dominik comment \thedominikcounter: #1 (line \the\inputlineno)}}

\newcounter{findingscounter}

\newcommand{\para}[1]{\vspace{2mm}\noindent\textbf{#1}}

\renewcommand\footnotetextcopyrightpermission[1]{} %

\begin{document}

\title{Mitigating Confirmation Bias on Twitter by Recommending Opposing Views}

\author{Elisabeth Lex}
\affiliation{%
  \institution{Graz University of Technology \& Know-Center GmbH, Austria}
}
\email{elisabeth.lex@tugraz.at}

\author{Mario Wagner}
\affiliation{%
  \institution{Graz University of Technology, Austria}
}
\email{Mario-wagner@gmx.at}

\author{Dominik Kowald}
\affiliation{%
  \institution{Know-Center GmbH,\\Austria}
}
\email{dkowald@know-center.at}

\begin{abstract}
Contemporary politics are often discussed on social media and thus, social media is an important information and news source \cite{Conover2011}. Politicians and parties increasingly use social media to spread their ideas, to interact with voters and/or critics and to shape the political debate \cite{Hong2016,Jungherr2016}. However, over the past years, social networks such as Facebook and Twitter have been accused of facilitating the spread of falsehoods, hatred and conspiracy theories \cite{economist2017}. Social networks have also been blamed to create so-called \emph{echo chambers} of like-minded people who are exposed mainly to views that are similar to their existing beliefs \cite{LevOn2009,Garimella2017}. Interestingly, people tend to perceive information so it confirms their existing beliefs, a phenomenon called \emph{confirmation bias} \cite{Silverman2017}. One countermeasure to confirmation bias is to give more prominence to opposing views \cite{LevOn2009}. 

\para{The Present Work.} In this work, we propose a content-based recommendation approach to increase exposure to opposing beliefs and opinions. Our aim is to help provide users with more diverse viewpoints on issues, which are discussed in partisan groups from different perspectives. Since due to the \emph{backfire effect} \cite{Silverman2017}, people's original beliefs tend to strengthen when challenged with counter evidence, we need to expose them to opposing viewpoints at the right time. The preliminary work presented here describes our first step into this direction. 

As illustrative showcase, we take the political debate on Twitter around the presidency of Donald Trump. In the 2016 US election, Twitter, among other social networks, played a strong role in the parties' campaigns and in shaping the political debate. For example, the campaign motto \emph{``Make America Great Again''} of President Trump has been translated in the widely adopted hashtag \emph{\#MAGA}. Quickly, people from all parts of the political spectrum have adopted the hashtag to discuss issues such as e.g., border policies from a variety of viewpoints. Pro-Trump Twitter users have used \emph{\#MAGA} to e.g., express their wish to build a wall on the border to Mexico in combination with \emph{\#BuildThatWall} while contra-Trump users have used \emph{\#MAGA} combined with \emph{\#NoWall} to express their opposing viewpoint on this issue. In other words, people discuss the same issues but from diverse perspectives. 

\para{Approach and Dataset.}
We base our work on Graells et al. \cite{Graells2013}, who created data portraits of users to determine a user's stance to the issue \emph{abortion in Chile}. They then used these data portraits to connect people of opposing views. For our work, we crawled a dataset consisting of two partisan groups with two stances: (i) pro-Trump users and (ii) contra-Trump users. To identify those two types of users, we used a set of manually selected hashtags. We selected these hashtags according to their discriminatory power to distinguish between pro and contra-Trump accounts as well as according to their reach\footnote{We assessed them with the service hashtag.org: \url{http://www.hashtags.org}.}. In case of pro-Trump accounts, we used the following hashtags: 'maga','tcot'\footnote{'tcot' stands for 'top conservatives on Twitter'.}, 'americafirst', 'trumptrain', 'presidenttrump', 'draintheswamp', 'fakenews', 'potus','buildthewall', 'presidentelecttrump'. 
In case of contra-Trump accounts, we used the hashtags: 'impeachtrump', 'theresistance', 'nobannowall', 'resist', 'trumprussia', 'impeach45', 'nottheenemy', 'resistance', notmypresident', 'iamamuslimtoo', 'nobannowallnoraids', 'fakepresident', 'dumptrump', 'trumplies'.

This procedure resulted in a dataset of 73,868 tweets from 39,698 accounts. Next, we extracted single group accounts, i.e., accounts that strictly use hashtags from only one group -- either the pro-Trump or the contra-Trump group since many hashtags appear in both groups. In order to further filter out managed accounts such as news channels, we created boxplots for the number of followers, the number of followees, the number of likes and the number of status updates. All accounts, which were below the first quartile and above the third quartile were filtered out, resulting in 6,913 accounts. We also removed accounts with a non-English user language, resulting in 5,672 accounts. We then downloaded the last 1,000 tweets from these remaining accounts, imported them into Apache Solr\footnote{\url{https://lucene.apache.org/solr/}}, normalized the texts to lowercase and performed tokenization, as well as stopword removal using the Snowball framework\footnote{\url{https://github.com/snowballstem}}. 

Since our initial classification is affected by the multiple meanings of a hashtag (e.g., \emph{\#MAGA}), we followed the method of \cite{Graells2013} to reliably classify the users into pro-Trump and contra-Trump users.
Thus, we created pro-Trump and contra-Trump issue stance vectors by concatenating all tweets of the users from each stance and extracting trigrams using TF-IDF. This results in our two issue stance vectors.
We did the same to create user stances, i.e., we concatenated the tweets of each user $u$ and ran TF-IDF to extract user $u$'s trigrams, resulting in a set of 5,672 user vectors. By calculating the cosine similarity between the issue stance vectors and each user vector, we can classify our users into one of the two issue stances. In total, we identified 2,150 pro-Trump users with 2,615,140 pro-Trump tweets and 3,522 contra-Trump users with 3,852,895 contra-Trump tweets.

To determine a user's personal preference about the stance, we extract the 15 most common trigrams from her tweets, as shown exemplary for a Twitter account in Figure \ref{fig:userex}.

\begin{figure}
  \centering
      \includegraphics[width=0.4\textwidth]{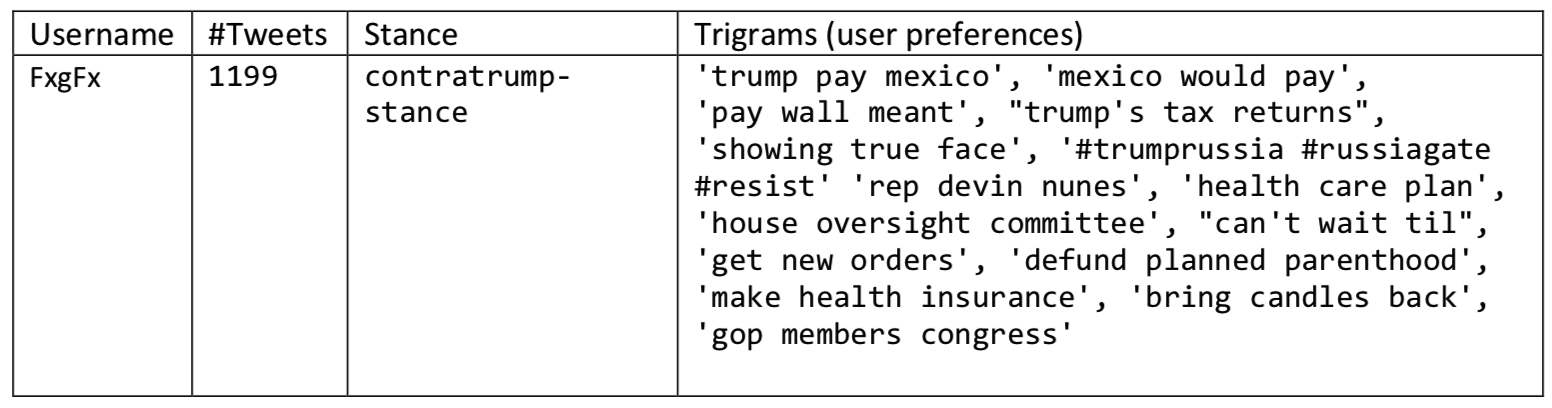}
  \caption{Example Twitter account ``FxgFx'', who has been classified into the contra-Trump stance and her most common trigrams, which serve as proxy for her preferences.\vspace{-3mm}}
  \label{fig:userex}
\end{figure}

We then implemented a content-based filtering recommendation approach by exploiting Apache Solr's MoreLikeThis functionality\footnote{We will share the link to our GitHub repository shortly.}. For each recommendation, we created a candidate set of 100k tweets, i.e., 50k random pro-Trump tweets and 50k random contra-Trump tweets from our dataset. Then, we computed the cosine similarity between a user's top trigram and the candidate set. An example recommendation set for the Twitter account ``FxgFx'' is shown in Figure \ref{fig:userexrec} for her top trigram ``trump pay mexico''. 

\begin{figure}
  \centering
      \includegraphics[width=0.4\textwidth]{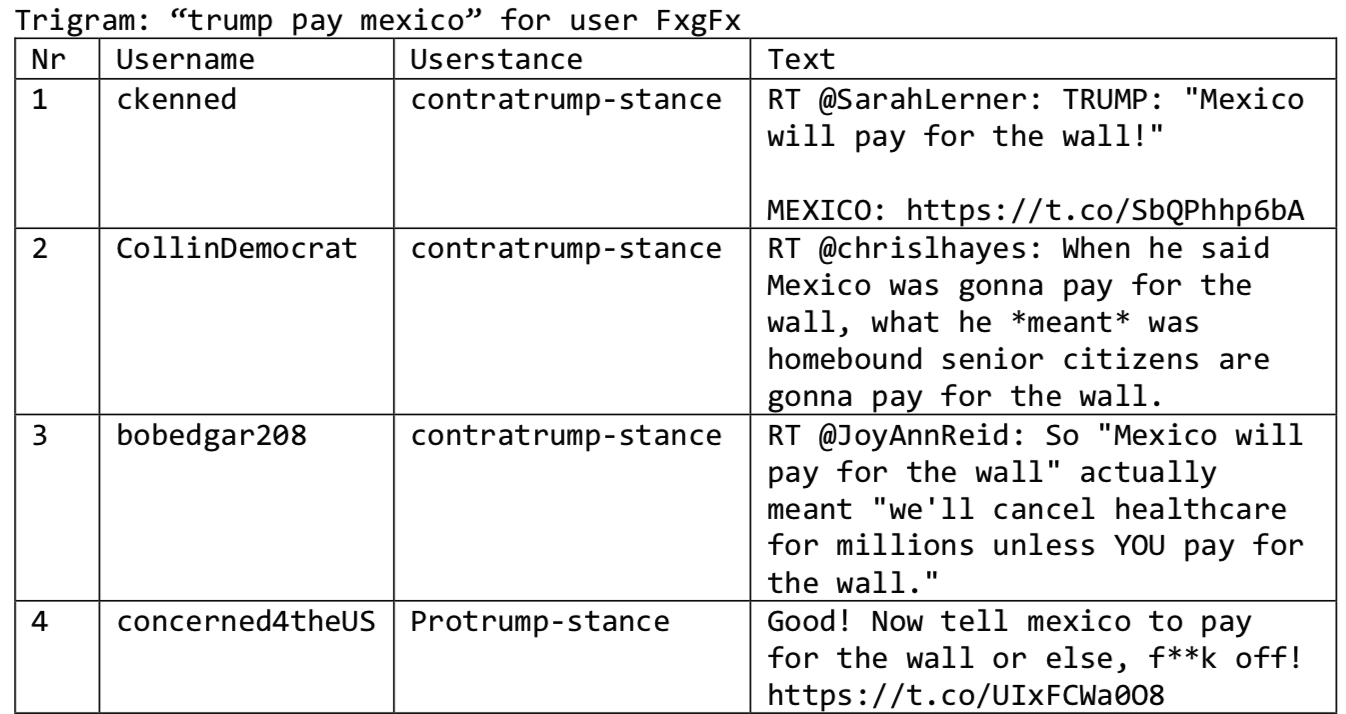}
  \caption{Ex.: Recommendations for Twitter account ``FxgFx''.\vspace{-3mm}}
  \label{fig:userexrec}
\end{figure}

\para{Evaluation.} Since we are interested in mitigating confirmation bias by increasing exposure to opposing views, we created four variants of our content-based filtering recommender: (i) ``Standard'', which recommends the 10 best matching tweets regardless of the user stance (as shown in Figure \ref{fig:userexrec}), (ii) ``Pro-Trump'', which only recommends pro-Trump tweets, (iii) ``Contra-Trump'', which only recommends contra-Trump tweets, and (iv) ``Hybrid'', which recommends half pro-Trump and half contra-Trump tweets (i.e., $5 + 5$). 

We evaluated these variants with beyond accuracy metrics of recommender systems research, i.e., we computed diversity and serendipity of recommendations \cite{kaminskas2017diversity}. Diversity is measured by the intra-list similarity metric, which sums all pairwise cosine similarities of the items in a given set and calculates the average of the sum. If a set has many similar items, the score is high, if the items are very different, the score is low. Serendipity \cite{Ge2010} measures how surprising the recommendations for a user are. In other words, serendipity denotes the distance between recommended items and their expected content. 

To further understand how diverse the pro-Trump and contra-Trump users in our dataset are per se, we computed the average topic similarity per user stance. We define the average topic similarity per issue stance as the average pairwise cosine similarity between all users of an issue stance (i.e., pro-Trump and contra-Trump). 

\para{Results and Discussion.} The results of our evaluation are given in Table \ref{tab:results} calculated for 1,500 randomly chosen pro-Trump users and 1,500 randomly chosen contra-Trump users. For the serendipity metric, as expected, the best results are achieved by the recommendation variant, which recommends tweets from the opposing view. This means that contra-Trump tweets provide the highest serendipity effect for pro-Trump users and vice versa. If both pro-Trump and contra-Trump tweets are mixed, serendipity gets lower.  
With respect to diversity, in the pro-Trump setting, the best results are achieved by the hybrid variant, which is also the behavior that we expected. However, in the contra-trump setting, we observe a different and rather surprising behavior since the best results here are achieved by the ``Pro-Trump'' variant of our recommender approach and not the hybrid one. One reason for this could be in the higher average topic similarity of contra-Trump users in our dataset. Interestingly, there is a big difference between both groups since the contra-Trump accounts exhibit an average topic similarity of 44.6\% while the pro-Trump accounts exhibit an average topic similarity of only 27.7\%. If the user group, in this case contra-Trump, has a high average topic similarity its inherent diversity is lower. Thus, diversity becomes lower if many tweets from a low diversity group are mixed into the recommendations. Consequently, better diversity results could be achieved if fewer (less than 50\%) of the more similar contra-Trump tweets and more of the diverse pro-Trump tweets are recommended. How to best engineer this ratio will be part of our future work.

\begin{table}[t!]
	\small
  \setlength{\tabcolsep}{5.0pt}	
  \centering	
    \begin{tabular}{l|l|c|c}
    \specialrule{.2em}{.1em}{.1em}
Issue stance 	  & Recommendation variant	  & Serendipity & Diversity \\\hline
Pro-Trump & Standard				   & .935 		   & .560		\\
		  & Pro-Trump				 & .943		   & .630		\\  
		  & Contra-Trump			& \textbf{.951} 		   & .695		\\   
		  & Hybrid	  			 	   & .946		   & \textbf{.728}		\\\hline
Contra-Trump & Standard				   & .924 		   & .441		\\
		  & Pro-Trump			   & \textbf{.957}		   & \textbf{.728}		\\  
		  & Contra-Trump		 & .925 		   & .487		\\
		  & Hybrid	  			   & .940		   & .701		\\
	\specialrule{.2em}{.1em}{.1em}								
    \end{tabular}
	 \caption{Our evaluation results with respect to recommendation serendipity and diversity. Here, the ``Hybrid'' approach recommends half pro-Trump and half contra-Trump tweets. \vspace{-8mm}}
  \label{tab:results}
\end{table}

\para{Conclusion and Future Work.}
We propose a hybrid content-based recommendation approach to mitigate confirmation bias and to help increase exposure to opposing views and beliefs. Our idea is to combine recommendations from partisan groups, i.e., pro-Trump and contra-Trump, into a mixed set that contains tweets from both sides. For evaluation, we turned to beyond-accuracy metrics of recommender systems, i.e., diversity and serendipity and we found that our approach lets us boost both metrics. In the longer run, we think of the problem at hand as an optimization task to achieve a trade-off between recommendation accuracy, serendipity and diversity.

For future work, we will research on the optimal mixing strategy to create the hybrid recommendations from pro-Trump and contra-Trump users. We also will study communication patterns in our dataset to better understand partisan and cross-partisan interactions. Also, we will research when to best add cross-partisan recommendations to increase the chance of recommendation acceptance \cite{Garimella2017}. One idea is to detect shifts in attention and user focus e.g., with the SUSTAIN \cite{Kopeinik2017} or ACT-R algorithm \cite{Kowald2017}, or shifts in conversational strength or sentiment. Finally, we also plan to verify our findings in larger Twitter data samples and with different political topics.

\para{Acknowledgments.} This work is funded by the Know-Center GmbH (Austrian COMET program) and the H2020 project AFEL (grant agreement: 687916).

\para{Keywords.} Confirmation Bias; Tweet Recommendations; Diversity; Serendipity; Polarization; Hybrid Recommendations

\end{abstract}

\maketitle

\balance

\begin{thebibliography}{00}


\ifx \showCODEN    \undefined \def \showCODEN     #1{\unskip}     \fi
\ifx \showDOI      \undefined \def \showDOI       #1{{\tt DOI:}\penalty0{#1}\ }
  \fi
\ifx \showISBNx    \undefined \def \showISBNx     #1{\unskip}     \fi
\ifx \showISBNxiii \undefined \def \showISBNxiii  #1{\unskip}     \fi
\ifx \showISSN     \undefined \def \showISSN      #1{\unskip}     \fi
\ifx \showLCCN     \undefined \def \showLCCN      #1{\unskip}     \fi
\ifx \shownote     \undefined \def \shownote      #1{#1}          \fi
\ifx \showarticletitle \undefined \def \showarticletitle #1{#1}   \fi
\ifx \showURL      \undefined \def \showURL       #1{#1}          \fi
\providecommand\bibfield[2]{#2}
\providecommand\bibinfo[2]{#2}
\providecommand\natexlab[1]{#1}
\providecommand\showeprint[2][]{arXiv:#2}

\bibitem[\protect\citeauthoryear{Berkeley}{Berkeley}{2017}]%
        {economist2017}
\bibfield{author}{\bibinfo{person}{Jon Berkeley}.}
  \bibinfo{year}{2017}\natexlab{}.
\newblock \bibinfo{title}{Do social media threaten democracy?}
\newblock   (\bibinfo{date}{November} \bibinfo{year}{2017}).
\newblock
\showURL{%
\url{https://www.economist.com/leaders/2017/11/04/do-social-media-threaten-democracy}}
\newblock
\shownote{[Online; posted 04-November-2017].}


\bibitem[\protect\citeauthoryear{Conover, Ratkiewicz, Francisco, Goncalves,
  Menczer, and Flammini}{Conover et~al\mbox{.}}{2011}]%
        {Conover2011}
\bibfield{author}{\bibinfo{person}{Michael Conover}, \bibinfo{person}{Jacob
  Ratkiewicz}, \bibinfo{person}{Matthew Francisco}, \bibinfo{person}{Bruno
  Goncalves}, \bibinfo{person}{Filippo Menczer}, {and}
  \bibinfo{person}{Alessandro Flammini}.} \bibinfo{year}{2011}\natexlab{}.
\newblock \showarticletitle{Political Polarization on Twitter}.
\newblock  (\bibinfo{year}{2011}).
\newblock
\showURL{%
\url{https://www.aaai.org/ocs/index.php/ICWSM/ICWSM11/paper/view/2847}}


\bibitem[\protect\citeauthoryear{Garimella, De~Francisci~Morales, Gionis, and
  Mathioudakis}{Garimella et~al\mbox{.}}{2017}]%
        {Garimella2017}
\bibfield{author}{\bibinfo{person}{Kiran Garimella}, \bibinfo{person}{Gianmarco
  De~Francisci~Morales}, \bibinfo{person}{Aristides Gionis}, {and}
  \bibinfo{person}{Michael Mathioudakis}.} \bibinfo{year}{2017}\natexlab{}.
\newblock \showarticletitle{Reducing Controversy by Connecting Opposing Views}.
  In \bibinfo{booktitle}{{\em Proceedings of the Tenth ACM International
  Conference on Web Search and Data Mining}} {\em (\bibinfo{series}{WSDM
  '17})}. \bibinfo{publisher}{ACM}, \bibinfo{address}{New York, NY, USA},
  \bibinfo{pages}{81--90}.
\newblock
\showISBNx{978-1-4503-4675-7}
\showDOI{%
\url{http://dx.doi.org/10.1145/3018661.3018703}}


\bibitem[\protect\citeauthoryear{Ge, Delgado-battenfeld, and Jannach}{Ge
  et~al\mbox{.}}{2010}]%
        {Ge2010}
\bibfield{author}{\bibinfo{person}{Mouzhi Ge}, \bibinfo{person}{Carla
  Delgado-battenfeld}, {and} \bibinfo{person}{Dietmar Jannach}.}
  \bibinfo{year}{2010}\natexlab{}.
\newblock \showarticletitle{Beyond accuracy: evaluating recommender systems by
  coverage and serendipity}. In \bibinfo{booktitle}{{\em In RecSys ’10}}.
  \bibinfo{pages}{257}.
\newblock


\bibitem[\protect\citeauthoryear{{Graells-Garrido}, {Lalmas}, and
  {Quercia}}{{Graells-Garrido} et~al\mbox{.}}{2013}]%
        {Graells2013}
\bibfield{author}{\bibinfo{person}{E. {Graells-Garrido}}, \bibinfo{person}{M.
  {Lalmas}}, {and} \bibinfo{person}{D. {Quercia}}.}
  \bibinfo{year}{2013}\natexlab{}.
\newblock \showarticletitle{{Data Portraits: Connecting People of Opposing
  Views}}.
\newblock \bibinfo{journal}{{\em ArXiv e-prints\/}} (\bibinfo{date}{Nov.}
  \bibinfo{year}{2013}).
\newblock
\showeprint[arxiv]{cs.HC/1311.4658}


\bibitem[\protect\citeauthoryear{Hong and Kim}{Hong and Kim}{2016}]%
        {Hong2016}
\bibfield{author}{\bibinfo{person}{Sounman Hong} {and}
  \bibinfo{person}{Sun~Hyoung Kim}.} \bibinfo{year}{2016}\natexlab{}.
\newblock \showarticletitle{Political polarization on twitter: Implications for
  the use of social media in digital governments}.
\newblock \bibinfo{journal}{{\em Government Information Quarterly\/}}
  \bibinfo{volume}{33}, \bibinfo{number}{4} (\bibinfo{year}{2016}),
  \bibinfo{pages}{777 -- 782}.
\newblock
\showISSN{0740-624X}
\showDOI{%
\url{http://dx.doi.org/https://doi.org/10.1016/j.giq.2016.04.007}}


\bibitem[\protect\citeauthoryear{Jungherr}{Jungherr}{2016}]%
        {Jungherr2016}
\bibfield{author}{\bibinfo{person}{Andreas Jungherr}.}
  \bibinfo{year}{2016}\natexlab{}.
\newblock \showarticletitle{Twitter use in election campaigns: A systematic
  literature review}.
\newblock \bibinfo{journal}{{\em Journal of information technology \&
  politics\/}} \bibinfo{volume}{13}, \bibinfo{number}{1}
  (\bibinfo{year}{2016}), \bibinfo{pages}{72--91}.
\newblock


\bibitem[\protect\citeauthoryear{Kaminskas and Bridge}{Kaminskas and
  Bridge}{2017}]%
        {kaminskas2017diversity}
\bibfield{author}{\bibinfo{person}{Marius Kaminskas} {and}
  \bibinfo{person}{Derek Bridge}.} \bibinfo{year}{2017}\natexlab{}.
\newblock \showarticletitle{Diversity, serendipity, novelty, and coverage: a
  survey and empirical analysis of beyond-accuracy objectives in recommender
  systems}.
\newblock \bibinfo{journal}{{\em ACM Transactions on Interactive Intelligent
  Systems (TiiS)\/}} \bibinfo{volume}{7}, \bibinfo{number}{1}
  (\bibinfo{year}{2017}), \bibinfo{pages}{2}.
\newblock


\bibitem[\protect\citeauthoryear{Kopeinik, Kowald, Hasani-Mavriqi, and
  Lex}{Kopeinik et~al\mbox{.}}{2017}]%
        {Kopeinik2017}
\bibfield{author}{\bibinfo{person}{Simone Kopeinik}, \bibinfo{person}{Dominik
  Kowald}, \bibinfo{person}{Ilire Hasani-Mavriqi}, {and}
  \bibinfo{person}{Elisabeth Lex}.} \bibinfo{year}{2017}\natexlab{}.
\newblock \showarticletitle{Improving Collaborative Filtering Using a Cognitive
  Model of Human Category Learning}.
\newblock \bibinfo{journal}{{\em The Journal of Web Science\/}}
  \bibinfo{volume}{2}, \bibinfo{number}{4} (\bibinfo{year}{2017}),
  \bibinfo{pages}{45--61}.
\newblock
\showDOI{%
\url{http://dx.doi.org/10.1561/106.00000007}}


\bibitem[\protect\citeauthoryear{Kowald, Pujari, and Lex}{Kowald
  et~al\mbox{.}}{2017}]%
        {Kowald2017}
\bibfield{author}{\bibinfo{person}{Dominik Kowald},
  \bibinfo{person}{Subhash~Chandra Pujari}, {and} \bibinfo{person}{Elisabeth
  Lex}.} \bibinfo{year}{2017}\natexlab{}.
\newblock \showarticletitle{Temporal Effects on Hashtag Reuse in Twitter: A
  Cognitive-Inspired Hashtag Recommendation Approach}. In
  \bibinfo{booktitle}{{\em Proceedings of the 26th International Conference on
  World Wide Web}} {\em (\bibinfo{series}{WWW '17})}.
  \bibinfo{publisher}{International World Wide Web Conferences Steering
  Committee}, \bibinfo{address}{Republic and Canton of Geneva, Switzerland},
  \bibinfo{pages}{1401--1410}.
\newblock
\showISBNx{978-1-4503-4913-0}
\showDOI{%
\url{http://dx.doi.org/10.1145/3038912.3052605}}


\bibitem[\protect\citeauthoryear{Lev-On and Manin}{Lev-On and Manin}{2009}]%
        {LevOn2009}
\bibfield{author}{\bibinfo{person}{Azi Lev-On} {and} \bibinfo{person}{Bernard
  Manin}.} \bibinfo{year}{2009}\natexlab{}.
\newblock \showarticletitle{Happy Accidents: Deliberation and Online Exposure
  to Opposing Views}.
\newblock \bibinfo{journal}{{\em Online Deliberation: Design, Research and
  Practice, Todd Davies, Seeta Gangadharan, eds., Forthcoming.\/}}
  (\bibinfo{date}{October} \bibinfo{year}{2009}).
\newblock
\newblock
\shownote{Available at SSRN: https://ssrn.com/abstract=1481869.}


\bibitem[\protect\citeauthoryear{Silverman}{Silverman}{2017}]%
        {Silverman2017}
\bibfield{author}{\bibinfo{person}{Craig Silverman}.}
  \bibinfo{year}{2017}\natexlab{}.
\newblock \bibinfo{title}{The Backfire Effect}.
\newblock   (\bibinfo{date}{June} \bibinfo{year}{2017}).
\newblock
\showURL{%
\url{https://archives.cjr.org/behind_the_news/the_backfire_effect.php}}
\newblock
\shownote{[Online; posted 17-June-2011].}


\end{thebibliography}

\end{document}